\documentclass[twocolumn,showpacs,amsmath,amstex,amssymb,mathfonts,prl]{revtex4-1}
\usepackage{graphicx}

\usepackage{tipa}
\usepackage{bm}

\usepackage{amsmath}
\usepackage{amssymb}
\usepackage{amsthm}
\usepackage{amsfonts}

\begin{document}

\title{Dirac Cone Metric and the Origin of the Spin Connections in Monolayer Graphene}

\author{Bo Yang}
\affiliation{Complex Systems Group, Institute of High Performance Computing, A*STAR, Singapore, 138632.}

\date{\today}
\begin{abstract}
We show that the modulation of the hopping amplitudes in the honeycomb lattice of the monolayer graphene uniquely defines a metric which corresponds to the shape of the Dirac cone near the Dirac points. The spin connection derived from this effective metric field agrees exactly with the microscopic tight-binding Hamiltonian. The paradoxical mismatch between the quantum field theory (QFT) approach and the microscopic tight binding model in the literature is resolved. The effective metric as seen by the sublattice pseudospin is different from the real space metric as defined by the two-dimensional manifold of the graphene monolayer. All relevant terms of the effective gauge field from the microscopic model is calculated exactly for a unimodular effective metric.
\end{abstract}

\maketitle 

The massless Dirac spectrum of the monolayer graphene sheet leads to many interesting properties\cite{wallace, geim, cn} in the long wavelength limit, where the sublattice pseudospin structure of the continuous model is analogous to that of the real spin of massless Dirac fermions moving in a two-dimensional manifold. For condensed matter systems without Lorentz invariance, a free spin $1/2$ Dirac fermion moving in a curved space responds to the external electromagnetic field and the curvature of the space via both $U(1)$ and $SU(2)$ local gauge invariance: the spinor couples to both the electromagnetic vector potential and the spin connection. Given the sublattice spinor structure of the graphene effective Hamiltonian, one would expect analogous geometric description in the form of effective gauge fields generated by the external perturbation to the lattice structure of the graphene sheet\cite{guinea}. 

There have been extensive efforts in modeling the strain and ripples of the monolayer graphene sheet in the form of the effective gauge fields, both from a microscopic point of view\cite{geim2,doussal,vozmediano,novoselov,vitor} and from the quantum field theoretical(QFT) approach used in treating Dirac spinors moving in a curved space\cite{guinea,vozmediano1,juan,lambiase}. The Landau levels from the effective gauge fields were also observed experimentally\cite{levy}. With the QFT approach, it is argued that the metric from either the two-dimensional manifold of graphene sheet or from the in-plane strain field introduces a spin connection that couples to the sublattice pseudospin. While the Dirac spectrum is robust against such ripples and strain field as long as the two Dirac cones from inequivalent points in the Brillouin zone do not merge\cite{cn1}, the microscopic origins of such a spin connection is still not clear. This is understandable, since a real spin couples to the real space spin connection due to its transformation property under rotations in real space. The situation with a pseudospin travelling in a curved space is more subtle. Recent calculations show that the QFT approach agrees with the microscopic description only up to a model dependent parameter, with the lowest order approximation of the metric deformation\cite{vozmediano1}. One would expect the spin connection can be derived exactly from the microscopic theory when the appropriate metric for the sublattice pseudospin is used in the QFT description.

Thus it has not been not clear if the effective gauge field from the microscopic hopping parameter modulation can really be interpreted as some form of the spin connection that couples to the pseudospin of the massless Dirac fermions in graphene. In this Letter, we show that the answer is yes, and the spin connection that couples to the sublattice pseudospin should \emph{not} be derived directly from the metric of the two-dimensional manifold as defined by the lattice structure of the graphene sheet in the real space. Instead, it should be derived from the metric defined by the shape of the Dirac cone in the \emph{momentum space}. This effective metric can be calculated microscopically from the lattice model, when the hopping amplitudes of the nearest neighbours in the honeycomb lattice are modified either by the ripples or by the in-plane strains. Our results show that the spin connection is subleading in the long wavelength as expected. The microscopic calculation derives the \emph{exact} spin connection corresponding to the Dirac cone metric, up to any order of the metric deformation and \emph{without} any model dependent parameters. The spatial dependence of the fermi velocity, as well as terms in the effective gauge field independent of the Gaussian curvature of the Dirac cone metric, are also exactly calculated. 

\begin{figure}
\includegraphics[width=7.5cm]{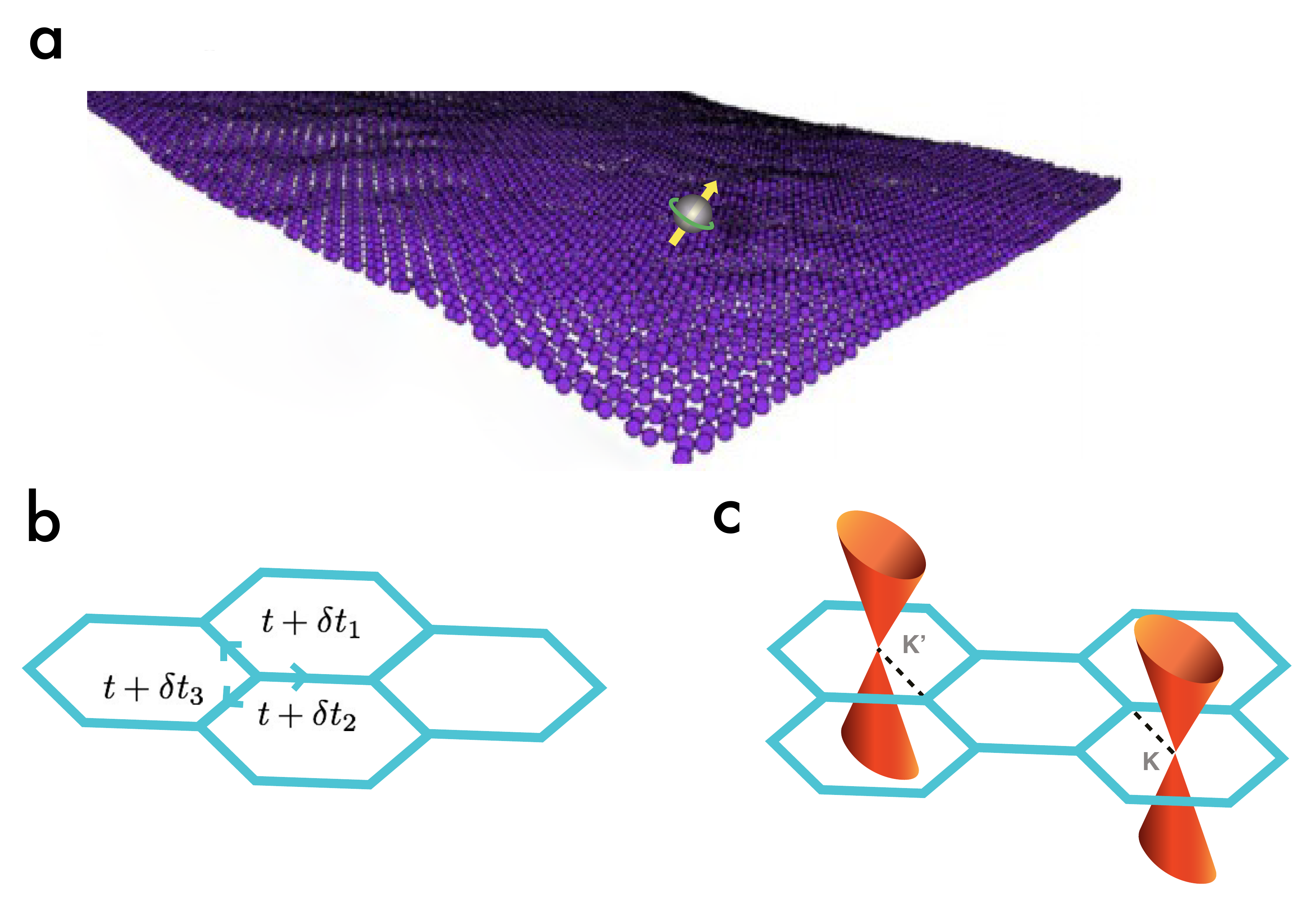}
\caption{(Color Online) a). The schematic drawing of a naturally occuring monolayer graphene sheet with ripples\cite{guinea}. The Dirac fermion is a $1/2$ pseudospin with the direction of the spin locked with respect to the direction of its momentum. The rotation of the pseudospin necessarily involves the rotation of its momentum vector. b). The honeycomb lattice in the real space, where the nearest neighbour hopping amplitudes are modified by $\delta t_1,\delta t_2,\delta t_3$ in different directions. c). The two physical effects of the hopping amplitude modification are the shift of the Dirac cone in the momentum space and the deformation and dilation of the shape of the Dirac cone. The background honeycomb lattice is the Brillouin zone of the unperturbed graphene monolayer, which serves as the reference. The shift, dilation and deformation can be calculated from a uniform modification of the hopping amplitude, which is locally valid as long as the modulation of the hopping amplitude modification varies slowly across the real space. }
\label{dirac}
\end{figure}

We start with the most general tight-binding model with the nearest neighbour hopping between site $A$ and site $B$:
\small
\begin{eqnarray}\label{h}
\mathcal H =- \sum_{\vec x, i}\left(t+\delta t_{i,\vec x}\right)c^\dagger_A(\vec x)c_B(\vec x+\vec r_i)+h.c.
\end{eqnarray}
\normalsize
Here $\vec x$ is the vector of the Bravais lattice, and $t\sim 2.7\text{eV}$ is the hopping amplitude of the ideal isotropic honeycomb lattice, whereby $\delta t_{i,\vec x}$ is the modulation to the hopping amplitude that depends on $\vec x$ and the indices of the three nearest neighbour vectors given by
\footnotesize
\begin{eqnarray}\label{3vectors}
\vec r_1=\left(0,\frac{-1}{\sqrt 3}\right)a,\vec r_2=\left(\frac{1}{2},\frac{1}{2\sqrt 3}\right)a,r_3=\left(-\frac{1}{2},\frac{1}{2\sqrt 3}\right)a
\end{eqnarray}
\normalsize
where $a$ is the lattice constant. The two inequivalent Dirac points in the momentum space from the isotropic lattice is given by $\vec K,\vec K'=\left(\pm 4\pi/(3a),0\right)$. Without loss of generality we now focus only on the low-energy states around $\vec K$. One can transform Eq.(\ref{h}) into the momentum space and do the long wavelength expansion around $\vec K$, keeping only terms linear in $\vec k$, where $\vec k$ is the momentum measured from $\vec K$. Transforming the Hamiltonian back to the real space, one has $\widetilde{\mathcal H}=\sum_{\vec x}\Psi^\dagger(\vec x)h(\vec x)\Psi(\vec x)$, with $\Psi(\vec x)=\left(c_A(\vec x),c_B(\vec x)\right)^\top$. For $h(\vec x)$ it is actually useful to define
\begin{eqnarray}
s_{1}&=&\frac{1}{3t}\left(\delta t_{1,\vec x}+\delta t_{2,\vec x}+\delta t_{3,\vec x}\right)+1\label{s1}\\
s_{2}&=&\frac{1}{3t}\left(\delta t_{1,\vec x}-\frac{1}{2}\left(\delta t_{2,\vec x}+\delta t_{3,\vec x}\right)\right)\label{s2}\\
s_{3}&=&\frac{1}{2\sqrt 3t}\left(\delta t_{2,\vec x}-\delta t_{3,\vec x}\right)\label{s3}
\end{eqnarray} 
where the spatial dependence of $s_i$ is omitted to clean up the notations. In this new set of ``coordinates" we have $h(\vec x)=v_F\left(\begin{array}{ccc}
0 & Q(\vec x) \\
Q^*(\vec x) & 0\end{array}\right)$ with
\begin{eqnarray}
Q(\vec x)&=&\left(s_{1}-s_{2}+is_{3}\right)P_x+i\left(s_{1}+s_{2}-is_{3}\right)P_y\nonumber\\
&&+\frac{1}{2}\left(\partial_ys_{1}+\partial_ys_{2}+\partial_xs_{3}-i\partial_xs_{1}+i\partial_xs_{2}-i\partial_ys_{3}\right)\nonumber\\
&&-\frac{3t}{v_F}\left(s_{2}+is_{3}\right)+O(P_a^2)\label{q}
\end{eqnarray}
where $P_a=-i\hbar\frac{\partial}{\partial a}$ is the momentum operator, $a=x,y$. The Fermi velocity is $v_F=\frac{\sqrt 3}{2}at$ and $O(P_a^2)$ contains higher derivatives. The first lines of Eq.(\ref{q}) clearly defines a spatially dependent effective fermi velocity $\widetilde v_F(\vec x)$ and an effective unimodular metric $\widetilde g^{ab}(\vec x)$ with $\det \widetilde g = 1$ that defines the shape of the Dirac cone. The second line of Eq.(\ref{q}) comes from the requirement that $h(\vec x)$ is Hermitian, and the third line is the well-known vector potential that contains no derivatives, describing the shift of the Dirac points in the momentum space. With some straightforward algebra we obtain
\begin{eqnarray}
\widetilde v_F(\vec x)&=&v_Fg(\vec x)^{\frac{1}{4}}\label{nvf}\\
\widetilde g^{xx}(\vec x)&=&\left(s_1^2+s_2^2+s_3^2-2s_1s_2\right)g(\vec x)^{-\frac{1}{2}}\label{gxx}\\
\widetilde g^{yy}(\vec x)&=&\left(s_1^2+s_2^2+s_3^2+2s_1s_2\right)g(\vec x)^{-\frac{1}{2}}\label{gyy}\\
\widetilde g^{xy}(\vec x)&=&\widetilde g^{yx}(\vec x)=2s_1s_3g(\vec x)^{-\frac{1}{2}}\label{gxy}
\end{eqnarray}
where $g(\vec x)=\left(s_1^2-s_2^2-s_3^2\right)^2$ is the determinant of the metric of the Dirac cone. Incidentally, the coordination of the carbon atoms in a honeycomb lattice is \textit{three}, providing just enough degrees of freedom for the parametrization of a general metric in a two-dimensional manifold; the three degrees of freedom are dilatation, squeezing and rotation respectively(see Fig.\ref{dirac}).

If the modulation of the hopping amplitude comes from a strain field that changes the relative positions between carbon atoms, an additional parameter $\beta=|\partial\log t/\partial\log a|$ is introduced\cite{cn,cn1}, together with a strain tensor $u_{ab}=\left(\partial_au_b+\partial_bu_a+\partial_ah\partial_bh\right)/2$, where $u_a$ are the in-plane displacement from the equilibrium point, and $h$ is the out-of-plane displacement\cite{vozmediano2}. The strain tensor defines the metric of the graphene manifold in the real space $g^s_{ab}=\eta_{ab}+2u_{ab}$, where $\eta_{ab}$ is the identity matrix corresponding to a flat and isotropic metric. This metric is, however, \emph{different} from the Dirac cone metric in the momentum space. Here it suffices to ignore the inhomogeneity of the strain tensor, and the modulation of the hopping amplitude is given by $s_{1}=1-\beta\left(u_{xx}+u_{yy}\right)/6, s_{2}=\beta(u_{xx}-u_{yy})/12, s_{3}=-\beta u_{xy}/6$; thus in terms of the strain tensor the Dirac cone metric is given by
\begin{eqnarray}
g(\vec x)^{\frac{1}{2}}\widetilde g^{xx}&=&1-\frac{1}{2}\beta u_{xx}-\frac{1}{6}\beta u_{yy}+O(u_{ab}^2)\label{ggxx}\\
g(\vec x)^{\frac{1}{2}}\widetilde g^{yy}&=&1-\frac{1}{6}\beta u_{xx}-\frac{1}{2}\beta u_{yy}+O(u_{ab}^2)\label{ggyy}\\
g(\vec x)^{\frac{1}{2}}\widetilde g^{xy}&=&-\frac{1}{3}\beta u_{xy}+O(u_{ab}^2)\label{ggxy}
\end{eqnarray}
which is manifestedly different from $g^s_{ab}$, as the latter is insensitive to the underlying lattice structure (and in particular $\beta$). Thus when the two-dimensional manifold of the graphene sheet is defined by the surface geometry of the substrate on which the graphene sheet is rested, the effective metric seen by the sublattice pseudospin is actually different from that of the graphene sheet in the real space itself. 

We now show that $g(\vec x)^{\frac{1}{2}}\widetilde g^{ab}(\vec x)$ is the physically relevant metric for the pseudospin, instead of $g^s_{ab}(\vec x)$ as defined from the strain tensor. The unimodular metric $\widetilde g^{ab}$ is expressed in terms of the complex vector $\widetilde\omega^a$, with $\widetilde g^{ab}(\vec x)=\left(\widetilde\omega^{a*}\widetilde\omega^b+\widetilde\omega^a\widetilde\omega^{b*}\right)/2$, and $\widetilde\omega^a=-i\epsilon^{ab}\widetilde\omega_b, \widetilde\omega^{a*}\widetilde\omega_a/2=1$. Here the Einstein summation convention is implied, $\epsilon^{ab}$ is the antisymmetrization tensor, and $a=x,y$ is the spatial index. Both the spin connection $\Omega_a$ and the Gaussian curvature $\mathcal K$ can be compactly expressed as\cite{haldane}
\begin{eqnarray}
\Omega_a(\vec x)&=&\epsilon^{bc}\widetilde\omega_c^*\partial_a\widetilde\omega_b-\frac{1}{2}\epsilon^{bc}\partial_b\widetilde g_{ac}\label{sc}\\
\mathcal K(\vec x)&=&\epsilon^{ab}\partial_a\Omega_b(\vec x)\nonumber\\
&=&\frac{1}{8}\epsilon^{ef}\epsilon_{ac}\widetilde g_{bd}\left(\partial_e\widetilde g_{ab}\right)\left(\partial_f\widetilde g_{cd}\right)-\frac{1}{2}\partial_a\partial_b\widetilde g^{ab}\label{k}
\end{eqnarray}
To compare Eq.(\ref{sc}) with the effective gauge field from the microscopic model in Eq.(\ref{q}), we just need to rewrite $h(\vec x)$ with Eq.(\ref{q}) to make the minimal coupling of the effective gauge field explicit: 
\begin{eqnarray}
&&h(\vec x)=\widetilde v_F\left(\begin{array}{ccc}
0 & \widetilde\omega^a(\vec x)\left(P_a+\mathcal A_a(\vec x)\right)\\
c.c& 0\end{array}\right)\label{tsh}
\end{eqnarray}
Here $c.c$ is the complex conjugate. The algebra is straightforward and we obtain the main result of this paper:
\begin{eqnarray}\label{gf}
\mathcal A_a(\vec{x})=\frac{2\sqrt 3}{a}\mathcal A^0_a(\vec x)+\frac{1}{2}\Omega_a(\vec x)
\end{eqnarray}
The effective gauge field in Eq.(\ref{gf}) thus has two parts. The first term is the well-known vector potential that leads to a pseudo-magnetic field. The factor of $1/2$ in front of the spin connection in the second term reflects the coupling of the $1/2$ sublattice pseudospin to the Gaussian curvature of the metric field as defined by the shape of the Dirac cone; it is thus the property of the quasiparticles at the Fermi surface. 

To analytically show that Eq.(\ref{gf}) can be derived from the microscopic model, it is sufficient to outline the derivation here for the special case where $g(\vec x)=1$. Note Eq.(\ref{tsh}) and Eq.(\ref{gf}), thus the main result of this work, holds for the general case when there \emph{is} a dilation of the Dirac cone metric, which modifies the Fermi velocity. The special case with the constraint $s_{1}^2=s_{2}^2+s_{3}^2+1$ only serves to make the derivation simpler; generalization to a general Dirac cone metric is straightforward though rather tedious.  By taking $\widetilde v_F=v_F$ in Eq.(\ref{gxx})$\sim$Eq.(\ref{gxy}), the second term in Eq.(\ref{sc}) can be easily calculated from Eq.(\ref{q}). For the first term in Eq.(\ref{sc}), one way to calculate it is to write $\widetilde\omega_x=\cosh\theta e^{i\phi}+\sinh\theta e^{-i\phi},\widetilde\omega_y=i\cosh\theta e^{i\phi}-i\sinh\theta e^{-i\phi}$, where $\theta$ and $\phi$ parametrizes the squeezing and rotation of the metric $\widetilde g^{ab}$. This gives $\cosh 2\theta=1+2\left(s_2^2+s_3^2\right), \sinh 2\theta\cos 2\phi=2s_1s_2,\sinh 2\theta\sin 2\phi=2s_1s_3$. Explicitly the spin connection is given by
\begin{eqnarray}\label{spinconnection}
\Omega_x(\vec x)&=&s_2\partial_xs_3-s_3\partial_xs_2+s_1\left(\partial_xs_3+\partial_ys_2\right)\nonumber\\
&&+\frac{s_3}{s_1}\left(s_2\partial_xs_2+s_3\partial_xs_3\right)\nonumber\\
&&+\left(2+\frac{s_2}{s_1}\right)\left(s_2\partial_ys_2+s_3\partial_ys_3\right)\\
\Omega_y(\vec x)&=&s_2\partial_ys_3-s_3\partial_ys_2+s_1\left(\partial_xs_2-\partial_ys_3\right)\nonumber\\
&&+\left(\frac{s_2}{s_1}-2\right)\left(s_2\partial_xs_2+s_3\partial_xs_3\right)\nonumber\\
&&-\frac{s_3}{s_1}\left(s_2\partial_ys_2+s_3\partial_ys_3\right)
\end{eqnarray}
One can also show that the leading effect of the pseudo-vector potential is explicitly given by
\begin{eqnarray}
\mathcal A^0_x(\vec x)&=&s_{3}^2-s_{2}^2-s_1s_2\label{sax}\\
\mathcal A^0_y(\vec x)&=&2s_{2}s_{3}-s_1s_3\label{say}
\end{eqnarray}

In the long wavelength limit where $|k|\ll 1/a$, the spin connection is subleading, and the effective gauge field is reduced to $\mathcal A^0_a(\vec x)$. Note in the linear approximation $s_1\sim 1$ and Eq.(\ref{sax}) and Eq.(\ref{say}) are reduced to the previously calculated gauge field induced by the strain. The terms higher in powers of $s_i$ reflects the correction to the shift of the position of the Dirac cone due to the variation of the metric in the momentum space. We would like to emphasize here Eq.(\ref{gf})$\sim$Eq.(\ref{say}) are exact for any modulation of the hopping amplitude, as long as the massless Dirac cones are robust.

With time-resersal symmetry, the effective metric around the two valleys $\vec K'$ and $\vec K$ are identical, but the effective gauge field will reserve its sign. For an ideal honeycomb lattice the eigenstates at energy $\mathcal E=\pm v_F|k|$ are given by $\frac{1}{\sqrt{2}}\left(\mp e^{i\phi_k/2},e^{-i\phi_k/2}\right)$ around $\vec K$ and $\frac{1}{\sqrt{2}}\left(\pm e^{-i\phi_k/2},e^{i\phi_k/2}\right)$ around $\vec K'$, with $\tan\phi_k =k_y/k_x$. It is well-known that the helicity in graphene is a good quantum number\cite{cn}. The helicity operators at $\vec K$ and $\vec K'$ are paired by time-reversal symmetry with $k_x\rightarrow k_x, k_y\rightarrow -k_y$. At the same energy, the time-reversal symmetry does not reverse the direction of the pseudospins around $\vec K$ and $\vec K'$ (in contrast to the real spin).  However, it does reverse the helicity, which has eigenvalues $\pm 1/2$ around the two valleys. It is thus more intuitive to treat the helicity eigenvalue as the coupling constant between the spinor and the spin connection.

The modulation of the nearest neighbour hopping amplitudes thus generates an effective metric field (Eq.(\ref{nvf})$\sim$Eq.(\ref{gxy})) relating to the shape of the Dirac cone, and a pseudo-magnetic field (Eq.(\ref{sax})$\sim$Eq.(\ref{say})) relating to the shift of the Dirac point in the momentum space. We take these two fields to be the same at both the valleys. There are two species of massless Dirac fermions in the graphene sheet, one from each valley. The Dirac fermions from valley $\vec K$ has helicity $+1/2$ and a ``valley charge" $2\sqrt 3/a$ that couples to the spin connection and the pseudo-magnetic field respectively. The Dirac fermions from valley $\vec K'$ is the time-reversal partner of those from $\vec K$, with helicity $-1/2$ and a ``valley charge" of $-2\sqrt 3/a$. For graphene sheets with topological defects such as pentagons and heptagons, these defects can be treated as sources of positive (pentagons) or negative (heptagons) gravity-like curvatures\cite{guinea}, which also couples to the helicities of the Dirac fermions.

In the abstract ``crystal frame" which is defined by the equilibrium position of the atoms, the hopping amplitudes captures completely the physics of the in-plane strain, height variation of the graphene sheet due to rippling, as well as other geometric features like wrinkles and conical singularities\cite{vitor4,kim}. It was also noted first in\cite{vitor2} and later in\cite{ vozmediano2}, that when graphene is probed by an external source, the ``lab frame" has to be used, where additional corrections arise from the deviation of the carbon atoms from their equilibrium position. While no additional pseudo-magnetic field is generated in the lab frame, both the fermi velocity and the effective metric is modified. However, one would expect the physics of the graphene sheets to be completely defined by the hopping between the lattice sites; a numerical diagonalization of the graphene systems does not need the information of the real space positions of the carbon atoms. Thus while this effect may have experimental ramifications, conceptually the physics remain the same. We defer detailed analysis elsewhere, so as not to over-complicate the message in this Letter. 

\emph{Conclusion and discussion}. we show that the treatment of the quasiparticles as $1/2$ pseudospins in monolayer graphene in the QFT covariant approach is exact, if we identify the metric of the curved space perceived by the pseudospin as the metric defined by the shape of the Dirac cone, and the latter depends on the microscopic details of the hopping amplitudes between the nearest neighbours. This metric is manifestedly the property of the quasiparticles at the fermi surface. The modification of the hopping amplitude not only induces such an effective metric in the momentum space, but also induces a pseudo-magnetic field that couples to the ``valley charge" of the quasiparticles. This pseudo-magnetic field shifts the two inequivalent Dirac cones in opposite directions due to time-reversal symmetry. The exact expression of both the spin connection and the pseudo-magnetic field are given in terms of the hopping amplitudes, and thus can be easily calculated. The two species of fermions from two the valleys at the fermi surface carry opposite ``valley charge" and helicity. When an external magnetic field is applied, the electric charge will be the third coupling constant. The interplay between the ``valley charge" and the electric charge when both the pseudo-magnetic field and the external magnetic field are present has been studied in\cite{vitor5,roy,roy2,roy3}; with a well-defined spin connection the transport of the sublattice pseudospin in a general two-dimensional manifold with or without broken time reversal symmetry can be fully characterized.

The shape of the Dirac cone in the momentum space, and thus the effective Dirac cone metric, can be easily measured experimentally when a uniform modification of the hopping amplitude over the entire graphene sheet is realized. On the other hand, the experimental measurement of the effect of the spin connection is generally difficult. This is because of the subleading nature of the spin connection; any modification in the transport of the Dirac fermions is dominated by the leading pseudo-magnetic field. Progress in the technologies of the artificial honeycomb lattice\cite{hari}, however, opens up the possibilities of much more flexible modulation of the hopping amplitudes on the microscopic scale, by the proper engineering of the positioning of the lattice sites. One can thus engineer the modulation such that the leading pseudo-magnetic field vanishes, while the pseudospin coupling between the Dirac fermion and the Gaussian curvature of the Dirac cone metric remains non-zero. One should note that all the effects calculated in this work is only valid in the long wavelength limit when the modulation of each of the three hopping amplitudes over space is small; nevertheless given the artificial lattice with vanishing pseudo-magnetic field, one can verify that the spin connection calculated from the Dirac cone metric instead of the real space metric should agree better with the experimental measurement.

\begin{acknowledgements}
I would like to thank Vitor Pereira from Graphene Research Center of the National University of Singapore, Bitan Roy from University of Maryland and F.D.M. Haldane from Princeton University for many helpful discussions.
\end{acknowledgements}

\end{document}